\documentclass[lettersize,journal]{IEEEtran}

\PassOptionsToPackage{hyphens}{url}
\usepackage{setspace}

\usepackage{amsmath}
\usepackage{amsthm}
\usepackage{amssymb}    
\newtheorem{theorem}{Theorem}
\newtheorem{property}{Property}

\usepackage{makecell}
\usepackage{longtable}
\usepackage{booktabs}
\usepackage{float}
\usepackage[table]{xcolor}   
\usepackage{multirow}
\usepackage{stfloats}

\usepackage[ruled,vlined]{algorithm2e}  

\usepackage{graphicx}
\usepackage{subfig}
\usepackage{url}
\usepackage{cite}
\usepackage{textcomp}

\usepackage{flushend}
\usepackage[colorlinks,linkcolor=red,anchorcolor=green,citecolor=blue]{hyperref}
\usepackage{cleveref}

\usepackage{bbding}
\usepackage{verbatim}
\crefname{figure}{fig}{figures}
\Crefname{figure}{Fig}{Figures}
\begin{document}

\title{A Weighted Byzantine Fault Tolerance Consensus Driven Trusted Multiple Large Language Models Network}

\author{Haoxiang Luo, Gang Sun,~\IEEEmembership{Senior Member,~IEEE}, Yinqiu Liu,   Dongcheng Zhao, \\ Dusit Niyato,~\IEEEmembership{Fellow,~IEEE}, Hongfang Yu,~\IEEEmembership{Senior Member,~IEEE}, Schahram Dustdar,~\IEEEmembership{Fellow,~IEEE}
\thanks{This work was supported by the Major Key Project of PCL under Grant PCL2024A05.}
\thanks{H. Luo, G. Sun, and H. Yu are with the School of Information and Communication Engineering, University of Electronic Science and Technology of China, Chengdu 611731, China (e-mail: lhx991115@163.com; \{gangsun, yuhf\} @uestc.edu.cn).}
\thanks{Y. Liu and D. Niyato are with the College of Computing and Data Science, Nanyang Technological University, Singapore 639798 (e-mail: yinqiu001@e.ntu.edu.sg; dniyato@ntu.edu.sg). }
\thanks{D. Zhao is with the Pengcheng Laboratory, Shenzhen 518055, China (e-mail:zhaodc11@gmail.com).}
\thanks{S. Dustdar is with the Distributed Systems Group, TU Wien, Vienna 1040, Austria, and also with the ICREA, Universitat Pompeu Fabra, Barcelona 08002, Spain (e-mail: dustdar@dsg.tuwien.ac.at).} \thanks{The corresponding author: Gang Sun.}}



\maketitle

\begin{abstract}
Large Language Models (LLMs) have achieved remarkable success across a wide range of applications. However, individual LLMs often produce inconsistent, biased, or hallucinated outputs due to limitations in their training corpora and model architectures. Recently, collaborative frameworks such as the Multi-LLM Network (MultiLLMN) have been introduced, enabling multiple LLMs to interact and jointly respond to user queries. Nevertheless, MultiLLMN architectures raise critical concerns regarding the reliability and security of the generated content, particularly in open environments where malicious or compromised LLMs may be present. Moreover, reliance on centralized coordination undermines system efficiency and introduces single points of failure.
In this paper, we propose a novel Trusted MultiLLMN framework, driven by a Weighted Byzantine Fault Tolerance (WBFT) blockchain consensus mechanism, to ensure the reliability, security, and efficiency of multi-LLM collaboration. In WBFT, voting weights are adaptively assigned to each LLM based on its response quality and trustworthiness, incentivizing reliable behavior, and reducing the impact of malicious nodes. Extensive simulations demonstrate that WBFT significantly improves both consensus security and efficiency compared to classical and modern consensus mechanisms, particularly under wireless network conditions. Furthermore, our evaluations reveal that Trusted MultiLLMN supported by WBFT can deliver higher-quality and more credible responses than both single LLMs and conventional MultiLLMNs, thereby providing a promising path toward building robust, decentralized AI collaboration networks.
\end{abstract}


\section{Introduction} \label{sec-I}

\IEEEPARstart{L}{arge} language models (LLMs) have become a cornerstone of AI, showing great capabilities in natural language understanding and generation \cite{hu2024federated}, \cite{chang2024survey}. These LLMs, represented by ChatGPT, Deepseek, and Claude, have been widely adopted in all aspects of society, such as education, healthcare, and information technology \cite{ray2023chatgpt}, \cite{jiang2024large}, \cite{wang2025generative}. 

With the deepening of LLM applications, a large number of LLMs developed by different commercial, academic, or educational organizations have emerged. 
However, considering the diverse learning corpus, model architectures, and training goals, the answers of different LLMs to the same query will naturally differ \cite{zhou2024larger}, \cite{gibney2025best}. 
Additionally, a single LLM is difficult to adapt to various application scenarios with heterogeneous contexts and requirements \cite{ding2024easy2hard}, \cite{lu2024merge}. 
Some LLMs even suffer from limitations and obsolescence in the training data, resulting in biased content generation and the ability to produce low-confidence outputs or hallucinations \cite{owens2024multi}, \cite{feng2024don}, \cite{wang2025wireless}. 
To address such challenges, the collaboration of multiple LLMs is on the agenda.
For instance, Wang et al. \cite{wang2025performance} leveraged GPT-3, GPT-4o, Llama3-8B, Llama3-Chinese, Doubao, SparkDesk, Qwen, and Kimi to jointly provide care programs for the elderly. 
To accelerate collaboration and communication between multiple LLMs, Marro et al. \cite{marro2024scalable} designed and developed a versatile, efficient, and portable communication protocol for them. This kind of collaborative LLM network is named as Multi-LLMs Network (MultiLLMN).

\subsection{Research Motivations}
Despite realizing inter-LLM collaborations, MultiLLMN causes several security concerns.
First, any User Equipment (UE) can access historical queries and answers. 
Consequently, privacy disclosure \cite{luo2024escm} and security attacks can frequently occur, especially in applications with private data such as medical and health information. 
In addition, malicious LLMs in MultiLLMN, such as WormGPT \cite{firdhous2023wormgpt}, can not only mislead UEs, but also escalate cybersecurity threats. 
Even ostensibly honest LLMs may run on compromised devices, resulting in tampered or dishonest responses. 
Furthermore, in the MultiLLMN architecture, we typically need to employ an authoritative third party to determine the most reliable answer from the responses of multiple LLMs.
The introduction of centralized authorities undermines the response efficiency of MultiLLMN and can cause single-point failures, hindering the deployment of MultiLLMN in latency-sensitive scenarios, such as autonomous driving. 
These concerns pose a significant challenge to the efficient collaboration of MultiLLMN and prevent UE from obtaining credible answers. 
Consequently, we present the following research questions.
\begin{itemize} 
\item \textbf{Q1}: How can users ensure that they receive the best and trusted response from MultiLLMN in an open network environment and with potentially malicious LLMs?
\item \textbf{Q2}: How can MultiLLMN eliminate single points of failure and achieve efficient response aggregation through multi-LLM collaboration?
\end{itemize}

To answer these questions, we leverage blockchain to ensure the reliability and security of information transmission in MultiLLMNs. 
Specifically, blockchain consensus mechanisms enable each LLM to make decisions independently of third-party authorities through a decentralized voting process \cite{luo2024convergence}, \cite{luowireless}. 
Existing research on blockchain-enabled LLMs has focused mainly on distributed training processes and the traceability of generated content, such as \cite{luo2024bc4llm}, \cite{zuo2024large}, and \cite{geren2025blockchain}. 
However, the design of a consensus-driven MultiLLMN remains largely unexplored. 
Most of the existing consensus, such as Practical Byzantine Fault Tolerance (PBFT) \cite{castro1999practical} and HotStuff \cite{yin2019hotstuff}, assign uniform trust values and equal voting weights to participants. 
We notice that this assumption deviates from real-world scenarios, where LLMs vary significantly in their response quality, generation capabilities, and susceptibility to malicious behavior.
Consequently, applying equal voting rights across all LLMs often undermines the ability of MultiLLMN to produce the highest-quality and most trustworthy responses for UEs. 
Hence, the following research question is raised.
\begin{itemize} 
\item \textbf{Q3}: How to assign a fair and reasonable consensus voting weight to each LLM by evaluating its trustworthiness and generation ability comprehensively, to assist MultiLLMN to generate the best quality response?
\end{itemize}

As a result, it is necessary to design a Weighted Byzantine Fault Tolerance (WBFT) consensus for MultiLLMNs to reduce the voting weight of LLMs with poor response quality, weak generation ability, and possibly malicious behavior, to obtain LLM answers more accurately.

\subsection{Our Contributions}
To the best of our knowledge, this is the first work to present a blockchain consensus for MultiLLMNs. Specifically, the contributions of this paper can be summarized as follows:
\begin{itemize} 
\item To avoid the limitations of a single LLM, we propose the innovative concept of MultiLLMN. Through inter-LLM collaborations, MultiLLMN mitigates biased responses and hallucinations arising from the deficiencies of individual models. 
\item To prevent malicious LLM behaviors and enhance the efficiency of MultiLLMNs in serving UEs, we introduce a blockchain consensus mechanism to securely drive inter-LLM collaboration. Additionally, we design an improved clustering-based optimization method that dynamically adapts the MultiLLMN network structure, facilitating the formation of a Trusted MultiLLMN and further strengthening its reliability and effectiveness.
\item To facilitate reliable response selection in MultiLLMN, we develop the WBFT consensus mechanism, which assigns fair and adaptive voting weights to each LLM based on their response quality and trustworthiness. This weighting strategy improves the reliability and robustness of the consensus process.  
\item Through extensive simulations, we demonstrate that WBFT outperforms traditional consensus mechanisms in both security and efficiency. Furthermore, Trusted MultiLLMN generates responses of higher quality and credibility compared to those from a single LLM or a MultiLLMN without consensus participation.
\end{itemize}

\subsection{Structure of This Paper}
The remainder of this paper is organized as follows. Section \ref{sec-II} reviews the related work. Section \ref{sec-III} describes how blockchain drives Trusted MultiLLMN to work. Then, we present the workflow of Trusted MultiLLMN enabled by blockchain in Section \ref{sec-III}. In Section \ref{sec-IV}, we demonstrate the design of WBFT. Then, we analyze the security and complexity of WBFT in Section \ref{sec-V}. Furthermore, we conduct extensive performance simulations of this framework in \ref{sec-VI}, which have validated its effectiveness and superiority in serving UE. Finally, we summarize this work in \ref{sec-VII}.

\section{Related Works} \label{sec-II}
In this section, we investigate the relevant work on inter-LLM collaboration and blockchain-enabled LLMs. Particularly, we compare existing research with our proposal in TABLE \ref{tab1} to illustrate our contributions.
\begin{table*}[!t]
\centering %
    \centering
    \caption{Scheme comparison}
    
    \renewcommand{\arrayrulewidth}{0.8pt} 
    \renewcommand{\tabcolsep}{10pt} 
    
    {\fontsize{8}{10}\selectfont 
     
    \begin{tabular}{|m{0.6cm}|m{7.5cm}|m{7.5cm}|} 
        \hline
         \textbf{Ref.} & \textbf{Contributions} & \textbf{Possible limitations}\\ 
        \hline
        
       \cite{owens2024multi} & Design a decentralized and centralized multi-LLM network to avoid the biased generated content by a single LLM & Lack of security and efficiency issues in the multi-LLM interaction\\ 
        \hline
        \cite{feng2024don} & Propose a competitive approach to determine highly reliable responses among multi-LLM in addition to cooperation & Lack of security and efficiency issues in the multi-LLM interaction\\ 
        \hline
        \cite{wang2025performance} &  Utilize multi-LLM to jointly provide services for elderly care& Lack of security and efficiency issues in the multi-LLM interaction\\ 
        \hline
        \cite{luo2024bc4llm} &   Explore the technical routes for credibility of LLM learning corpora, training processes, and generated content by applying blockchain & Do not overcome the single LLM limitations, such as generated content bias and hallucinations \\ 
        \hline
        \cite{feng2025one} & Organize multi-LLM into a hierarchical structure based on levels of access and information exchange for efficient collaboration. & Lack of security issues in the multi-LLM interaction\\ 
         \hline
         \cite{shen2024learning} & Construct a collaboration method between the general basic model and the dedicated LLM& Lack of security and efficiency issues in the multi-LLM interaction\\ 
           \hline
          \cite{wan2024knowledge} &Enhance the reasoning ability of a single LLM through the knowledge fusion from multi-LLM & Lack of security issues in the multi-LLM interaction\\ 
           \hline
        \cite{zuo2024federated} & Empower the security of the LLM distributed training process with blockchain & Do not overcome the capacity limitations of a single LLM \\ 
            \hline
        \cite{chen2024blockagents} & Utilize blockchain to achieve unified and secure collaboration among multiple agents in LLM & Do not overcome the capacity limitations of a single LLM \\ 
            \hline
        \cite{liu2024blockchain} & Empower the traceability and immutability of AIGC content with blockchain & Do not overcome the capacity limitations of a single LLM \\ 
            \hline
        \cite{bouchiha2024llmchain} & Develop a reputation system based on blockchain to evaluate the generated content credibility 
        from LLM & Do not overcome the capacity limitations of a single LLM\\ 
            \hline
        \cite{lin2024blockchain} & Utilize blockchain to achieve efficient and trustworthy AIGC services in Metaverse & Do not overcome the capacity limitations of a single LLM\\ 
            \hline
            
        \cellcolor{cyan} Our work & \cellcolor{cyan} Design the WBFT blockchain consensus-driven Multi-LLM collaboration to provide high-quality responses for UE & \cellcolor{cyan} Add additional blockchain costs and interaction delays between LLMs  \\ 
        \hline
        
    \end{tabular}}
    
    \label{tab1}
\end{table*}

\subsection{Collaboration among Multiple LLMs}

A single LLM often struggles to comprehensively address real-world requirements, suffering from issues such as biased outputs, low-confidence generations, and hallucinations due to limitations in its training data. To overcome these inherent weaknesses, the inter-LLM collaboration has emerged as a promising direction and attracted great research attention.

For instance, Feng et al. \cite{feng2024don} designed a collaboration framework involving three LLMs to mitigate the knowledge gaps of a single LLM caused by outdated or insufficient training data. Their framework achieved a 19.3\% performance improvement across four tasks in different domains. 
Similarly, in \cite{feng2025one}, the authors argued that a single LLM cannot adequately represent the real-world data distribution or the diversity of human perspectives, and that this limitation cannot be resolved merely by training more powerful models.
Therefore, they proposed a hierarchical multi-LLM collaboration structure spanning API, text, and logical levels. 
In addition, Owens et al. \cite{owens2024multi} also investigated persistent bias in LLM outputs, originating from limited training data. 
Despite advances in bias mitigation techniques, such as data augmentation and fine-tuning, biased output remains a challenge. 
Consequently, they developed a multi-LLM communication model to reduce output bias.
Moreover, Shen et al. \cite{shen2024learning} introduced Co-LLM, a joint system where a general-purpose LLM invokes domain-specific expert models, achieving superior performance in instruction following, question answering, and reasoning tasks compared to standalone LLMs. 
Furthermore, knowledge distillation techniques have been employed to continuously transfer and consolidate knowledge from multiple LLMs into a target model, enhancing its generative capacity \cite{wan2024knowledge}. 
A recent practical application of multi-LLM collaboration appears in elderly care \cite{wang2025performance}. 
The authors employed eight mainstream LLMs to collectively provide seniors with services such as electronic payments, daily living assistance, recreational support, security alerts, and emotional companionship. This work highlights the strong potential of multi-LLM collaboration in real-world service scenarios.

However, existing work only considers the cooperation of multiple LLMs and cannot guarantee the trustworthiness of LLM-generated content. Moreover, how multiple LLMs can work together to provide the best response needs to be studied.

\subsection{Blockchain-enabled LLM}
Due to the existence of malicious LLMs and attackers in MultiLLMN, the research on trustworthy LLMs has been put on the agenda. This vision often relies on blockchain technology that integrates security features such as decentralization, traceability, and immutability.

For example, to enrich LLM training datasets, Zuo et al. \cite{zuo2024federated} developed a blockchain-based federated learning framework that enables various private databases to securely contribute training data.
To protect LLM training processes from Byzantine behaviors, Chen et al. \cite{chen2024blockagents} proposed BlockAgents, which integrates blockchain into LLM training to resist adversarial threats. 
Their framework reduces the impact of poisoning attacks on model accuracy to less than 3\% and the success rate of backdoor attacks to below 5\%, demonstrating strong robustness. 
Additionally, Liu et al. \cite{liu2024blockchain} leveraged blockchain to provide trusted endorsement and protection for AI-Generated Content (AIGC) products, offering traceable verification of ownership changes through smart contracts and incentive mechanisms to promote free circulation. 
Furthermore, Bouchiha et al. \cite{bouchiha2024llmchain} developed LLMChain, a blockchain-based reputation system designed to evaluate and monitor LLM behavior, addressing vulnerabilities such as hallucinations, unreliable reasoning, and harmful content generation. 
Similarly, Lin et al. \cite{lin2024blockchain} designed a blockchain smart contract-based verification mechanism to prevent random outcomes in AIGC services, thus improving service reliability in the Metaverse. 
Luo et al. \cite{luo2024bc4llm} systematically analyzed LLM trustworthiness from three key perspectives (i.e., learning corpus, training processes, and generated content) and emphasized the critical role of blockchain technologies across these dimensions.

Existing work has focused primarily on ensuring the trustworthiness of individual LLMs. However, research on the trustworthiness of multi-LLM collaboration remains limited. Ensuring trust in a collaborative environment is inherently more challenging, as it should account not only for the open network environment but also for malicious behaviors induced by compromised or adversarial LLMs.

\section{Blockchain-Driven Multiple-LLM Network} \label{sec-III}
In this section, we introduce the architecture of Trusted MultiLLMN driven by blockchains. The commonly used notations are summarized in TABLE \ref{tab2}.

\begin{table}[t]
\centering %
    \centering
    \caption{Key Notations}
    
    \renewcommand{\arrayrulewidth}{0.5pt} 
    \renewcommand{\tabcolsep}{10pt} 
    
    {\fontsize{8}{10}\selectfont 
    \begin{tabular}{|m{1.1cm}|m{6.2cm}|} 
        \hline
        \textbf{Notations} & \textbf{Definitions} \\ 
        \hline
         $A_{i,j}^r$ & The response quality weight assigned by the  $i$-th LLM to the $j$-th LLM in the $r$-th round consensus\\
        \hline
         $B_{i,j}^r$ & The trust weight assigned by the  $i$-th LLM to the $j$-th LLM in the $r$-th round consensus\\
         \hline
         $D_{i,h}^r$ & The encrypted data of the $h$-th response initiated by the $i$-th LLM in the $r$-th round consensus\\
        \hline
        $L_{i,j}^r$ & The communication latency for the $j$-th LLM to the $i$-th LLM in the $r$-th round consensus\\
        \hline
        $P_{i,h}^r$ & The proof of $D_{i,h}^r$\\
        \hline
        $Q_{i,j}^r$ & The response quality score of the $i$-th LLM to the $j$-th LLM in the $r$-th round consensus\\
         \hline
        $T_{i,j}^r$ & The trust score of the $i$-th LLM to the $j$-th LLM in the $r$-th round consensus\\
         \hline
        $V_{i,j}^r$ & The feature vectors of the $j$-th LLM for the $i$-th LLM in the $r$-th round consensus\\
        \hline
         $W_{i,j}^r$ & The weight assigned by the $i$-th LLM to the $j$-th LLM in the $r$-th round consensus\\
        \hline
         $b_{i,h}^r$ & The block added to the chain for the $h$-th response initiated by the $i$-th LLM in the $r$-th round consensus\\
        \hline
        $c_{j,h}^r$ & The confirmation message of the $j$-th LLM for the leader in the commit phase\\
        \hline
        $v_{j,h}^r$ & The voting value of the $j$-th LLM for $D_{i,h}^r$\\
          \hline
   \end{tabular}}
    
    \label{tab2}
\end{table}

\subsection{Workflow of Trusted MultiLLMN}
Regardless of their specific architectures, all LLMs possess the ability to generate content and assess the rationality of generated outputs. 
Therefore, we regard each LLM as one full node within the blockchain-driven network, participating directly in consensus processes and contributing to the validation of information exchanges. 
However, malicious LLMs may produce misleading content and provide dishonest evaluations of other LLMs' outputs, introducing vulnerabilities into the MultiLLMN. This challenge will be addressed through the implementation of our proposed consensus mechanism, which is detailed in the subsequent sections.

\begin{figure}[!t]
\centering
 \includegraphics[width=3.4in]{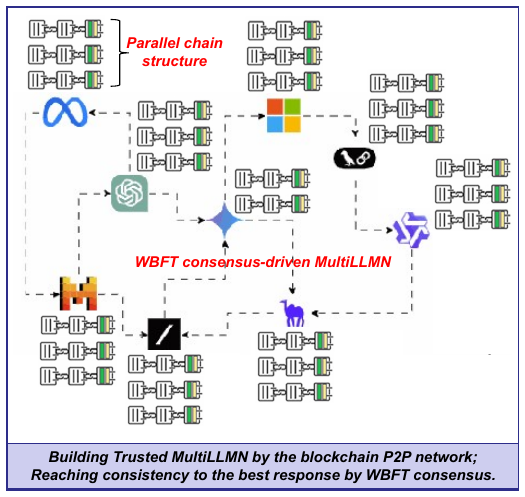}
   \caption{The blockchain-driven Trusted MultiLLMN. This network is built on a blockchain P2P network and has a decentralized architecture. The best response for UE is derived from the WBFT consensus. The consensus results are packaged into blocks and linked to the chains maintained by distributed LLMs.}
\label{fig1}
\end{figure}

Building upon mainstream LLMs, including Llama, WizardLM, GPT, and Gemini, we develop a blockchain-driven MultiLLMN, as illustrated in Fig.~\ref{fig1}. This network architecture not only enables collaborative interactions among heterogeneous LLMs but also inherently enhances credibility and trustworthiness through blockchain integration. The operational workflow is outlined as follows.

\emph{1) UE Requests:} The UE submits a response request to Trusted MultiLLMN.

\emph{2) LLM Answer Generation:} Each LLM generates a response to UE requests. These responses are subsequently disseminated among LLMs via broadcast protocols within the blockchain's Peer-to-Peer (P2P) network infrastructure.

\emph{3) Blockchain Consensus:} Consensus serves as a key technique in identifying the optimal response among multiple LLMs. In the proposed framework, a voting consensus mechanism is employed to evaluate and select the final proposal, which is then relayed back to the UE requirements by designated consensus leaders. The proposal is automatically determined by the WBFT consensus, which will be described in the next section. 

\emph{4) Block Package:} The optimal response, as determined by WBFT, is encapsulated within a block, which undergoes confirmation by each LLM. To guarantee the security, immutability, and traceability of the consensus outcomes, the block is meticulously designed to incorporate the hash value of the optimal solution alongside its precise timestamp.

\emph{5) Blockchain Extension:}  Blocks are correlated with respective parallel chains and stored in a distributed manner across smart devices. Specifically, each LLM extends and maintains the chain that incorporates its own responses, thereby enhancing the processing efficacy of the Trusted MultiLLMN. Note that each chain is documented by the host device of every LLM to preserve the decentralized character of the blockchain system.      

Through the above five steps, the Trusted MultiLLMN can be established, which delivers responses with blockchain-guaranteed security and reliability.

\subsection{Dynamic Optimization of Trusted MultiLLMN}
While the workflow of Trusted MultiLLMN ensures basic functionality, further optimization is needed to enhance response quality for UEs and improve the efficiency of blockchain consensus. To this end, we introduce a dynamic networking mechanism for Trusted MultiLLMN based on a clustering algorithm, named Hierarchical Secure Clustering (HSC). This approach not only adapts to variations in the network environment and LLM states but also significantly improves the scalability of Trusted MultiLLMN, enabling it to accommodate a larger number of LLMs.

HSC strategically categorizes all LLM nodes into two types: Core Cluster Nodes (CCNs) and Edge Cluster Nodes (ECNs). Only CCNs actively participate in the WBFT consensus mechanism, while ECNs receive consensus results through broadcast messages from their cluster's designated CCN. This hierarchical structure significantly reduces consensus latency and enhances the parallel distribution of consensus outcomes, thereby optimizing the Trusted MultiLLMN's response efficiency to UEs' requests.
 
For effective deployment in wireless networks, our Trusted MultiLLMN framework requires optimization to accommodate diverse operational conditions. The HSC algorithm therefore incorporates two critical parameters in the LLM feature vectors: trust value and communication latency. This approach effectively mitigates Byzantine LLMs' interference, ensuring consistency within the Trusted MultiLLMN, as represented by
\begin{equation}
V_{i,j}^r=[\omega T_{i,j}^r, (1-\omega)L_{i,j}^r].
\end{equation}
$V_{i,j}^r$ represents the feature vectors of the $j$-th LLM from the view of the $i$-th LLM in the $r$-th round of consensus. $T_{i,j}^r$ denotes the trust value assigned by $i$-th LLM to $j$-th LLM. $L_{i,j}^r$ denotes communication latency from $j$-th to $i$-th LLM. Finally, $\omega$ serves as the proportional regulator.

Then, the K-means++ algorithm \cite{li2022collaborative} is applied to determine the number of clusters $K$. Specifically, we add a penalty term $\lambda$ for dynamically controlling the number of clusters based on minimizing the sum of squares of the distance between the feature vectors of the nodes and the cluster center $\mu_k$ (i.e., the tightness of the cluster), i.e.,
\begin{equation}
\min \sum_{k=1}^K \sum_{V_{i,j}^r \in C_k} \lVert V_{i,j}^r - \mu_k \rVert^2 + \lambda.
\end{equation}
$C_k$ represents the $k$-th cluster divided by HSC. This equation can be solved using the Elbow method \cite{liu2020determine}. Next, we iteratively update the cluster centers until we identify exactly $K$ CCNs, with each CCN responsible for maintaining one distinct cluster, as expressed by:
\begin{equation}
\mu_{k}^{(t + 1)} = \frac{\sum_{j \in C_{k}}\left( \frac{T_{i,j}^{r}}{L_{i,j}^{r}} \right)^{\gamma} \cdot V_{i,j}^{r}}{\sum_{j \in C_{k}}\left( \frac{T_{i,j}^{r}}{L_{i,j}^{r}} \right)^{\gamma}},
\end{equation}
where $\gamma$ is a nonlinear regulatory factor. 
This equation allows HSC to select LLM nodes with high reputation and low latency as CCNs to optimize the cluster structure of MultiLLMN.
Furthermore, HSC can dynamically adapt to changing conditions in each consensus round for two key reasons. First, Byzantine LLMs exhibit inconsistent malicious behaviors, causing their trust values $T_{i,j}^{r}$ to fluctuate between rounds. Second, the inherent instability of wireless environments leads to temporal variations in communication latency $L_{i,j}^{r}$. These dynamic factors require HSC to continuously adjust cluster formations to maintain optimal performance and security throughout the consensus process.

\section{Weighted Byzantine Fault Tolerance Consensus} \label{sec-IV}
In this section, we illustrate the design of WBFT, including the weight allocation scheme and the consensus process.

\subsection{Weight Allocation Scheme}
In WBFT, the weight assigned to each LLM comprises two components: response quality, which reflects content generation capability, and trustworthiness, which indicates whether the LLM and its associated device exhibit malicious behavior. Any LLM can serve as a consensus initiator and thus assign variable weights to peer LLMs. Formally, we denote the weight assigned by the $i$-th LLM to the $j$-th LLM during the $r$-th round of consensus by $W_{i,j}^r$, which is defined as
\begin{equation}
\begin{aligned}
 \label{eq1}
& W_{i,j}^r = \alpha A_{i,j}^r + \beta B_{i,j}^r, \\
& \alpha + \beta = 1, \\
& \sum_{j=1}^n W_{i,j}^r = \sum_{j=1}^n A_{i,j}^r = \sum_{j=1}^n B_{i,j}^r = 1,
\end{aligned}
\end{equation}
\noindent where $A_{i,j}^r$ and $B_{i,j}^r$ denote the response quality and trust weights assigned by the $i$-th LLM to the $j$-th LLM in the $r$-th round consensus, respectively. The parameters $\alpha$ and $\beta$ determine the relative importance of these two weight components and can be adjusted according to specific requirements. $n$ denotes the total number of LLMs participating in the Trusted MultiLLMN.

In each round of consensus, the LLM receives both responses to UE queries and voting information from other LLMs. Consequently, at the beginning of each consensus round, the consensus initiator has the prerogative to dynamically re-calibrate the response quality and trust weights of other LLMs according to their performance in the last round.
Furthermore, weights $A_{i,j}^r$ and $B_{i,j}^r$ can be determined by the dual criteria of response quality and trustworthiness exhibited by all other LLMs, namely
\begin{equation}
\begin{aligned}
\label{eq2}
& A_{i,j}^r = \frac{Q_{i,j}^r}{\sum_{j=1}^n Q_{i,j}^r}, \quad \\
& B_{i,j}^r = \frac{T_{i,j}^r}{\sum_{j=1}^n T_{i,j}^r},
\end{aligned}
\end{equation}
where $Q_{i,j}^r$ and $T_{i,j}^r$ represent the answer quality and trustworthiness of the $i$-th LLM to the $j$-th LLM in the $r$-th round consensus, respectively.

\subsection{Consensus Design}
In a MultiLLMN, UEs can access the LLM functionality through any participant. We suppose that the LLM receiving UE's requests serves as the consensus leader, while other LLMs act as followers. Notably, due to the design of HSC, only $K$ CCNs participate in the consensus process, meaning all participating follower LLMs are CCNs.

To ensure secure query-response transmission within the MultiLLMN, we pre-assign two public and private key pairs to each LLM since each can function as either a consensus leader or follower. Specifically, these pairs are $(PK_i^L, SK_i^L)$ and $(PK_i^F, SK_i^F)$. The first pair is used exclusively when LLM $i$ serves as a leader, while the second pair is employed when it acts as a follower. Each LLM's public keys are shared network-wide for encrypting query-response data, while private keys remain secured for decryption operations.

\begin{figure*}[!t]
   \centering
  \includegraphics[width=7in]{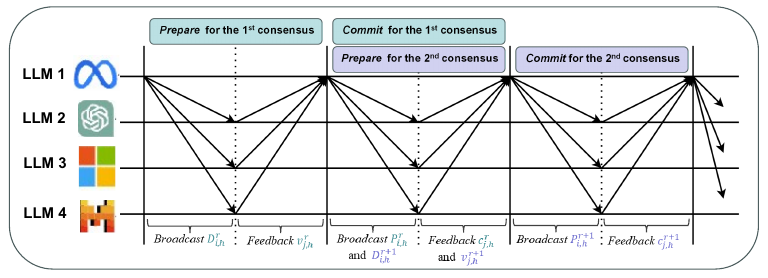}
   \caption{The consensus process of Weighted Byzantine Fault Tolerance (WBFT) with the pipeline mechanism. It allows the \emph{prepare} phase of the $(r+1)$-th round consensus to be initiated during the \emph{commit} phase of the $r$-th round consensus.}
   \label{fig2}
\end{figure*}

As illustrated in Fig. \ref{fig2}, WBFT consensus comprises two voting rounds, namely the \emph{prepare} phase (as shown in Algorithm \ref{alg1}) and the \emph{commit} phase (as shown in Algorithm \ref{alg2}). Below, we elaborate on the process of each phase.

\begin{itemize}

\begin{algorithm}[!t]
\caption{Prepare Phase}\label{alg1}
\SetKwInOut{Input}{Input}
\SetKwInOut{Output}{Output}

\underline{Leader Operations:}\\
1. Generate response: $R_L \leftarrow \text{generate\_response}(Q)$ \\
2. Encrypt data: $D_{i,h}^r \leftarrow \text{encrypt}(Q \parallel R_L, PK_j^F)$ \tcp*{h = current block height} 

3. Broadcast $D_{i,h}^r$ to all followers 
\BlankLine

\underline{Follower Operations:}\\
1. Decrypt data: $(Q, R_L) \leftarrow \text{decrypt}(D_{i,h}^r, SK_j^F)$ \\
2. Verify authenticity of $(Q, R_L)$ \\
3. \If{verification passes}{
    Generate response: $R_F \leftarrow \text{generate\_response}(Q)$ \\
    4. \eIf{$\text{quality}(R_F) > \text{quality}(R_L)$}{
        $v_{j,h}^r \leftarrow 0$ \tcp*{Follower's response is better}
    }{
        $v_{j,h}^r \leftarrow 1$ \tcp*{Leader's response is better}
    }
    5. Send $\text{encrypt}( v_{j,h}^r \parallel R_F, PK_i^L)$ to leader
}
\Else{
    Discard $D_{i,h}^r$ \tcp*{Invalid data detected}
}
\BlankLine
\underline{Leader Operations:}\\
1. Receive encrypted votes from followers\\
2. \If{ $\sum \text{weight}(v_{j,h}^r = 1) + \text{leader\_self\_weight}>2/3$}{
    Generate proof: $P_{i,h}^r \leftarrow \text{generate\_proof}(D_{i,h}^r)$  \\
}
\Else{
    Consensus termination.
}
\end{algorithm}

\begin{algorithm}[!t]
\caption{Commit Phase}\label{alg2}
\SetKwInOut{Input}{Input}
\SetKwInOut{Output}{Output}

\underline{Leader Operations:}\\
1. Broadcast $P_{i,h}^r$ to all followers 
\BlankLine

\underline{Follower Operations:}\\
1. Decrypt $P_{i,h}^r$ by threshold signature \\
2. Verify the Byzantine character of the leader \\
3. \If{verification passes}{
    expunge$P_{i,h}^{(r-1)}$ \\
    Transmit their confirmation messages back
to the leader \\
}
\Else{
    Discard $P_{i,h}^r$ \tcp*{Invalid data detected}
}
\BlankLine
\underline{Leader Operations:}\\
1. Receive confirmation messages from followers\\
2. \If{$\sum \text{weight}(c_{j,h}^r = 1) + \text{leader\_self\_weight}>2/3$}{
    Reach consistency: $R_L$ with its hash value and timestamp packaged as $b_{i,h}^r$ is appended to the chain.   \\
}
\Else{
    Consensus termination.
}
\end{algorithm}

\item \textbf{\emph{Prepare Phase:}} The consensus leader initially encrypts the UE's query $Q$ and its own generated response $R_L$ using the public key $PK_j^F$, subsequently broadcasting this encrypted data, denoted as $D_{i,h}^r$, to $n-1$ follower LLMs.  
Here, $h$ denotes the total number of responses initiated by the leader and concurrently represents the block height within the chain maintained by said leader.    

Each follower then employs their pre-assigned private key $SK_j^F$ to decrypt the received data and validate its authenticity.   Subsequently, each follower transmits a vote, $v_{j,h}^r$, back to the leader. Each vote comprises the vote outcome and the follower's own response $R_F$ to the UE's query $Q$ and is encrypted by the leader's public key $PK_i^L$. 
The voting mechanism dictates that a follower assigns a vote value of 0 if its generated response exceeds the leader's in terms of quality, and 1 otherwise.

The leader decrypts the voting information using the follower's public key $PK_j^F$ and tallies the voting values. 
When the cumulative weight of votes with value 1 exceeds two-thirds (including the leader's own vote weight), the leader generates a proof $P_{i,h}^r$ attesting that $D_{i,h}^r$ has been successfully verified. 
This proof is implemented as a threshold signature \cite{yin2019hotstuff} of the aggregated votes.

\item \textbf{\emph{Commit Phase:}} The consensus leader broadcasts proof $P_{i,h}^r$ to all followers, signaling that the response has been verified as optimal. Each follower authenticates this proof using threshold signature techniques to verify both the proof itself and the leader's non-Byzantine status. Upon successful validation, followers remove the previous proof $P_{i,h-1}^r$ to reduce storage requirements, then transmit confirmation messages $c_{j,h}^r$ to the leader. Once the leader receives confirmations representing at least a two-thirds weighted majority, the response $R_L$ with its hash value and timestamp is encapsulated in a block $b_{i,h}^r$ and appended to the corresponding chain.
In contrast, if the leader is detected exhibiting malicious behavior, the MultiLLMN system recommends that the UE obtain responses from LLMs on different devices to avoid potential misinformation from the compromised leader.
\end{itemize}

\begin{figure}[!t]
\centering
 \includegraphics[width=3.4in]{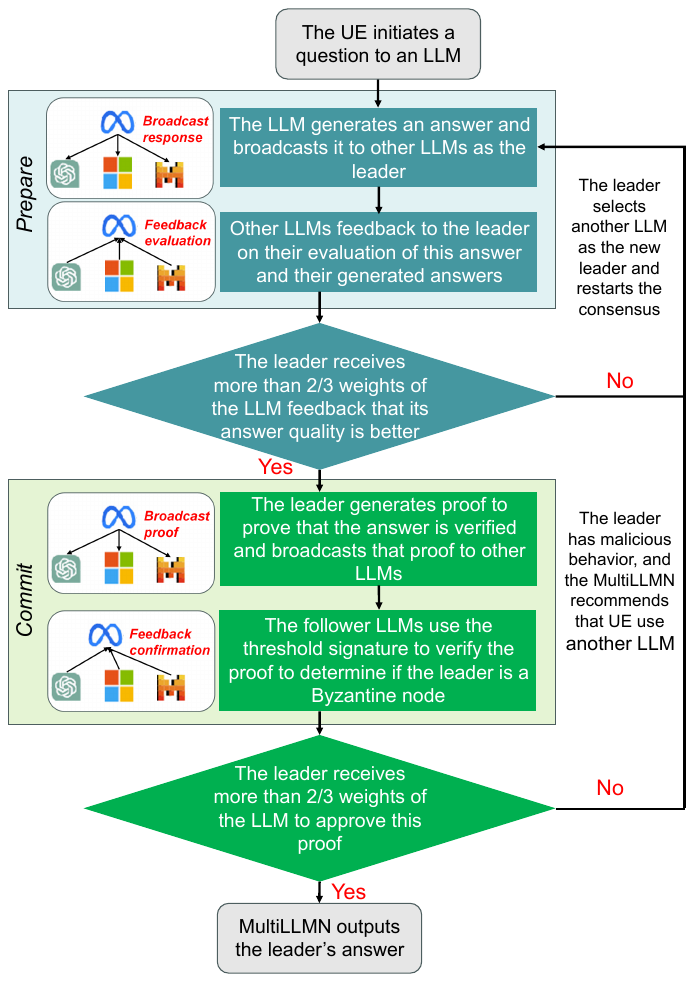}
   \caption{The \emph{prepare} and \emph{commit} phases of WBFT consensus.}
\label{fig3}
\end{figure}

Fig. \ref{fig2} illustrates the overall consensus process. Moreover, Fig. \ref{fig3} provides deeper insight into the dual voting stages. After completing these two rounds, the leader can evaluate the followers' response quality and trustworthiness based on the analysis of generated answers and voting feedback. This evaluation enables dynamic recalibration of voting weights for subsequent consensus rounds, highlighting a key advantage of LLMs, i.e., their ability to exercise independent judgment within the consensus mechanism.

Observing that our dual-round voting consensus mechanism naturally segments into distinct phases, we leverage this structural feature to implement a pipelined processing framework \cite{yu2024tierflow}, \cite{luo2024blockchain} to further improve efficiency. As shown in Fig. \ref{fig2}, while the commit phase for block $b_{i,h}^r$ is being executed, the prepare phase for the subsequent block $b_{i,h'}^r$ can be initiated concurrently, where $h'$ denotes the next block height in the blockchain managed by the $j$-th LLM. This pipeline architecture significantly reduces the waiting time between consecutive requests.

The optimal responses from independent consensus processes initiated by different leaders are assembled into separate blocks and added to chains maintained by their respective leaders, creating a parallel chain architecture. Although each LLM primarily manages its own chain, it also maintains copies of all other chains to ensure system-wide consistency and enable distributed storage of query-response across the MultiLLMN.

\section{Performance Analysis} \label{sec-V}
In this section, we analyze the consensus security, efficiency, and complexity of WBFT.

\subsection{Consensus Security}
Without loss of generality, we measure consensus security by the consensus success rate, as discussed in \cite{luo2023esia} and \cite{li2020scalable}. 
Specifically, we define $v_{j,h}^r = 1$ to indicate that a follower believes it cannot generate a response superior to that of the leader, whereas $v_{j,h}^r = 0$ implies the follower considers its own response to be of higher quality.  
Thus, successful consensus requires the following condition, denoted as Event $Y$: 
\begin{equation} 
Y = \sum_{j=1}^K W_{i,j}^r v_{j,h}^r > \frac{2}{3}. 
\end{equation}

Event $Y$ involves the summation of multiple products of random variables, making its direct computation highly complex. 
To address this challenge, we adopt two approaches to approximate the probability of Event $Y$. 
First, the Monte Carlo method provides a heuristic estimation based on random sampling. 
Second, when the number of LLMs, denoted by $n$, is sufficiently large, the Law of Large Numbers \cite{revesz2014laws} implies that the distribution of assigned LLM weights can be approximated by a normal distribution over the bounded interval $(0,1)$.
Consequently, the weights $W_{i,j}^r$ are assumed to follow a distribution $N_1\left(\mu, \sigma^2\right)$. 
Meanwhile, the voting outcomes $v_{j,h}^r$ adhere to a Bernoulli distribution. 
Since the assigned weights and the voting outcomes are independent, $W_{i,j}^r$ and $v_{j,h}^r$ are mutually independent random variables. 
By invoking the properties of linear combinations of independent random variables, the expected value and variance of Event $Y$ can be expressed as:
\begin{equation}
\begin{aligned}
E(Y) &= E\left(\sum_{j=1}^n W_{i,j}^r v_{j,h}^r\right) = \sum_{j=1}^n E\left(W_{i,j}^r v_{j,h}^r\right) \\
&= \sum_{j=1}^n E\left(W_{i,j}^r\right) E\left(v_{j,h}^r\right) = \sum_{j=1}^n \mu p = p,
\end{aligned}
\end{equation}
\begin{equation}
\begin{aligned}
D(Y) &= D\left(\sum_{j=1}^n W_{i,j}^r v_{j,h}^r\right) = \sum_{j=1}^n D\left(W_{i,j}^r v_{j,h}^r\right) \\
&= \sum_{j=1}^n \left[E\left(W_{i,j}^2\right)D\left(v_{j,h}^r\right) + E^2\left(W_{i,j}^r\right)D\left(v_{j,h}^r\right)\right] \\
&= \sum_{j=1}^n \left[p\sigma^2 + \mu^2 p(1-p)\right] = np\sigma^2 + \frac{1}{n}p(1-p),
\end{aligned}
\end{equation}
where $p$ represents the probability of $v_{j,h}^r=1$.

Following the Central Limit Theorem \cite{dudley2014uniform}, when the number of LLMs $n$ becomes sufficiently large, linear combinations of independent and identically distributed random variables approach a normal distribution. Consequently, the probability of Event $Y$ approximates a normal distribution with parameters $N_2\left(p, np\sigma^2 + \frac{1}{n}p(1-p)\right)$.
Therefore, the consensus security of WBFT, i.e.,  $P(Y > 2/3)$ can be expressed as
\begin{equation}
\begin{aligned}
p_{\text{WBFT}} &= 1 - P(Y \leq 2/3) \\ & = 1 - \Phi\left(\frac{2/3 - p}{\sqrt{np\sigma^2 + \frac{1}{n}p(1-p)}}\right),
\end{aligned}
\label{eq6}
\end{equation}
where $\Phi$ denotes the normal distribution $N(0,1)$. By restricting the voting rights of LLMs with low credibility and weak generation capabilities, WBFT will achieve higher consensus security.

\subsection{Consensus Efficiency}
We consider two metrics to reflect consensus efficiency, namely throughput and latency.

To facilitate the adaptable deployment and utilization of MultiLLMN, thereby augmenting its scope of application, we implement the proposed WBFT-driven Trusted MultiLLMN within wireless network contexts. According to \cite{chang2019optimizing}, \cite{yu2020low},\cite{luo2023symbiotic}, and \cite{cao2022v2v}, we can conclude that in wireless networks, the consensus latency is intrinsically linked to the consensus success rate, namely
\begin{equation}
1 - P_l = f_Q\left(\frac{NTBC - NTBR + \frac{\log NTB}{2}}{(\log e)\sqrt{NTB}}\right),
\end{equation}
where $P_l$ denotes the channel transmission success rate. Consequently, $1-P_l$ means the transmission failure rate associated with this channel. This failure probability can be generalized to the scenario where $v_{j,k}^r=0$, as it also implies that the leader has not received the follower's endorsement of its response. Furthermore, $f_Q$ denotes the Q-function. In this context, $T$ and $N$ represent the latency of a channel and the number of subcarriers, respectively, with $N$ set to 1 in this paper. Additionally, $B$, $R$, and $C$ represent bandwidth, transmission rate, and channel capacity, respectively.

Both the \emph{prepare} and \emph{commit} phases of the WBFT consensus involve two communication directions: \emph{downlink} transmissions, where the leader broadcasts messages, and \emph{uplink} transmissions, where followers send feedback. The communication structure of each phase is consistent with that of the Raft consensus algorithm \cite{luo2024energy}, \cite{luo2024symbiotic}, \cite{luo2023performance}. Accordingly, the latency associated with the WBFT consensus can be expressed as
\begin{equation}
t_c=2nT.
\end{equation}

For consensus throughput, it can be expressed as the number of UE requests processed by MultiLLMN per unit of time, whose unit is Transactions per Second (TPS). It can be said that, benefiting from the parallel pipeline processing of UE responses, WBFT has a higher throughput than traditional consensus.

\subsection{Correctness Analysis}
WBFT adopts a structure similar to PBFT, ensuring that as long as the number of faulty nodes does not exceed one-third of the total, the consensus satisfies three core properties: non-forking, consistency, and liveness. Assuming a total of $n = 3f + 1$ nodes, with at most $f$ Byzantine nodes, the corresponding proofs are presented as follows.
\begin{property}[Nonforking]
For any two valid committed blocks $b_{i_1,h_1}^{r_1}$ and $b_{i_2,h_2}^{r_2}$, if $i_1 = i_2$ then $h_1 = h_2$, where $b_{i,h}^{r}$ denotes the $h$-th block generated by $i$-th LLM in the $r$-th consensus round.
\end{property}

\noindent \emph{Proof.}
Suppose contradictory blocks $b_{i_1,h_1}^{r}$ and $b_{i_2,h_2}^{r}$ ($h_1 \neq h_2$) both receive $(2f+1)/n$ vote weights. With $n=3f+1$, we can acquire
\begin{equation}
2\cdot(2f+1)/n - (3f+1)/n = (f+1)/n.
\end{equation}
It implies $(f+1)/n$ vote weights for both blocks, violating the protocol rules. Hence, $h_1$ should equal $h_2$.
$\hfill\blacksquare$

\begin{property}[Consistency]
If the block $b_{i,h}^{r}$ is committed by any honest node, all honest nodes will eventually commit it.
\end{property}

\noindent \emph{Proof.}
Commitment requires proof $P_{i,h+1}^{r}$ to obtain $(2f+1)/n$ vote weights. Let $VW_{\mathsf{received}}$ denote the received affirmative vote weights ($VW_{\mathsf{received}} \geq (2f+1)/n$). At least $(2f+1)/n - f/n = (f+1)/n$ votes come from $VW_{\mathsf{received}}$. Since weights of Byzantine nodes $\leq f$, majority voting can ensure eventual global consistency.
$\hfill\blacksquare$

\begin{theorem}
For any response $R_L$ submitted by the honest nodes, there exists a finite time ${T_\mathsf{finite}}$ such that $R_L$ will be included in block $b_{i,j}^r$ generated by an honest leader.
\end{theorem}

\noindent \emph{Proof.}
 The liveness proof requires analyzing two mutually exclusive scenarios:

\emph{Case 1. Transaction Recovery Protocol:}
Suppose that leader $L_i^r$ proposes block $b_{i,h}^r$ containing $R_L$ with the following conditions:
\begin{enumerate}
    \item The follower node $F_j^r$ receives $b_{i,b}^r$ but lacks $R_L$.
    \item $F_j^r$ extracts the $R_L$ hash value $h(.)$ from $b_{i,h}^r$.
    \item By broadcasting $h(R_L)$, $F_j^r$ requests $R_L$ from neighbors via $\mathsf{GetData}(h(R_L))$ messages.
    \item The honest nodes that receive $R_L$ respond with $\mathsf{TxResponse}(R_L)$, validated through:
\begin{equation}
 \mathsf{VerifyTx}(h(R_L), \mathsf{sig}_{\mathsf{tx}}) = \mathsf{True}.
\end{equation}

\item After receiving $R_L$, $F_j^r$ executes:
\begin{equation}
    \mathsf{Vote}(b_{i,h}^r) \iff \mathsf{ValidateBlock}(b_{i,h}^r) = \mathsf{True}.
\end{equation}
    
\end{enumerate}
This process guarantees that all honest nodes can reconstruct $b_{i,h}^r$ in a bounded time $\Delta_1$.

\emph{Case 2. Voting Retry Protocol:}
If leader $L_i^r$ fails to receive more than $2/3$ of the vote weights in favor within the time window $T_{\mathsf{init}}$ due to network delays:
\begin{enumerate}
    \item When the network jitter, $VW_{\mathsf{received}} < (2f+1)/n$.
    \item Leader triggers timeout event:
\begin{equation}
    T_{\mathsf{retry}} = T_{\mathsf{init}} + \delta_{\mathsf{backoff}},
\end{equation}
      where $\delta_{\mathsf{backoff}}$ follows exponential increasing sequence.
    \item At $T_\mathsf{finite} = T_{\mathsf{retry}}$, leader reinitializes voting with new timestamp:
\begin{equation}
    b_{i,h'}^{r} \gets \mathsf{Repropose}(b_{i,h}^r, \mathsf{ts}_{\mathsf{new}}).
\end{equation}
    \item The process repeats until:
\begin{equation}
    \exists \, T_\mathsf{finite} \leq T_{\mathsf{max}}, \quad VW_{\mathsf{committed}} \geq (2f + 1)/n,
\end{equation}
    where $VW_{\mathsf{committed}}$ represents the affirmative vote weights of \emph{commit} phase.
\end{enumerate}
Therefore, the probability of success after $k$ times of attempts can be expressed as
\begin{equation}
p_{\mathsf{success}}(k) = 1 - \left(\frac{f}{3f + 1}\right)^k \quad \xrightarrow{k \to \infty} \quad 1.
\end{equation}
$\hfill\blacksquare$

\noindent \textbf{Corollary 1.} \emph{The proposed WBFT achieves liveness with probability $1$ under partial synchrony assumptions, where message delays are bounded by known time $\Delta_{\mathsf{network}}$.}

\subsection{Consensus Complexity}
Compared to the communication complexity of $O(n^2)$ exhibited by PBFT \cite{castro1999practical}, the proposed WBFT significantly reduces the communication complexity to $O(K)$. This improvement is achieved by restricting communication to interactions solely between the leader and the followers, thereby avoiding the extensive inter-follower consultations required during the two-round voting process \cite{luo2024multi}. In PBFT, follower-to-follower negotiations typically involve repeated flooding broadcasts, which substantially increase communication overhead.

Furthermore, WBFT maintains an advantage even over the $O(n)$ complexity of other parallel multi-chain consensus mechanisms, such as Vote as Proof (VaaP) \cite{fu2022votes}, \cite{luo2024efficient}. Specifically, WBFT not only narrows the scope of consensus participation through the HSC algorithm but also restricts the voting influence of malicious LLMs. By reducing the likelihood of leader re-elections, caused by difficulties in securing the required two-thirds voting weight, WBFT further lowers the overall consensus complexity.

\section{Performance Evaluation} \label{sec-VI}
In this section, we validate the efficiency of the proposed blockchain-driven Trusted MultiLLMN and WBFT.

\subsection{Parameter Acquisition and Setting}
In the first round of consensus in MultiLLMN, individual LLMs lack prior knowledge of the response quality and trustworthiness of their peers. As a result, it is difficult for LLMs to assign appropriate voting weights to one another.

To address this initial challenge of response quality assessment, we design a comprehensive evaluation framework. Specifically, we construct question sets spanning five distinct scenarios and engage a panel of 15 volunteers, carefully selected to ensure geographical diversity, to evaluate the content generation capabilities (i.e., response quality) of ten widely recognized LLMs. The volunteer distribution includes China (4 people), the United States (4 people), the United Kingdom (2 people), Singapore (2 people), Australia (1 person), the Netherlands (1 person), and Saudi Arabia (1 person).

The ten LLMs evaluated in our experiments: Llama 3.3, WizardLM 2, GPT-4o, Gemini 2 Flash, ERNIE Bot 4.0, SparkDesk V4.0, Qwen 2.5, Doubao Pro 4k, Hunyuan-Large, and Kimi. All LLMs are deployed on a high-performance server equipped with a 96-core Intel(R) Xeon(R) Gold 5220R CPU @ 2.20GHz and 1 TB of memory. The experimental scenarios are designed to cover a broad range of LLM competencies, including memory retention, everyday task performance, artistic creativity, logical reasoning, and code generation. The final category is specifically designed to support ongoing optimization in wireless network domains. The evaluation questions are as follows.

\begin{table*}[!t]
\centering %
    \centering
    \caption{LLMs Generation Capability Scores}
    \renewcommand{\arrayrulewidth}{0.5pt} 
    \renewcommand{\tabcolsep}{10pt} 
    
    {\fontsize{8}{10}\selectfont 
    \begin{tabular}{|m{1.75cm}<{\centering}|m{0.63cm}<{\centering}|m{1.21cm}<{\centering}|m{0.56cm}<{\centering}|m{0.93cm}<{\centering}|m{0.89cm}<{\centering}|m{1.22cm}<{\centering}|m{0.57cm}<{\centering}|m{0.81cm}<{\centering}|m{1cm}<{\centering}|m{0.56cm}<{\centering}|} 
        \hline
        \textbf{Volunteers order} & \textbf{Llama 3.3} & \textbf{WizardLM 2} & \textbf{GPT-4o} & \textbf{Geimini 2 Flash}& \textbf{ERNIE Bot 4.0} & \textbf{SparkDesk V4.0}& \textbf{Qwen 2.5} & \textbf{Doubao pro 4k} & \textbf{Hunyuan Large} & \textbf{Kimi}  \\ 
        \hline

        1 & 82 & 80& 77 &75 & 79 & 81& 82& 80& 79& 82\\
        \hline
        
        2 & 85 & 83& 84 &84 & 82 & 83& 84& 83& 83& 84\\
        \hline

         3 & 75 & 78& 85 & 76 & 78 & 72& 90& 85& 80& 93\\
        \hline

        4 & 70&	75&	80&	73	&75	&70	&86	&82	&78	&90\\
        \hline

        5 & 78	&77&	83&	75&	77&	73&	88	&84&	82&	82\\
        \hline

        6 & 72	&76	&81	&82	&76	&81&	85&	83	&79	&81\\
        \hline

        7 & 76&	79	&84	&84	&79&	74	&87&	75&	81&	83\\
        \hline

        8 & 73	&78&	82&	73&	77&	72	&86	&84	&78&	90\\
        \hline

        9 & 82&	74&	89&	78&	85&	79&	93&	78&	86&	86\\
        \hline

        10 & 80	&82&	87&	86&	83&	77	&91&	86&	84&	79\\
        \hline

        11 & 81	&83&	88&	77	&84	&78&	92&	77	&85&	85\\
        \hline

        12 & 83	&84&	89&	78&	86	&83&	93&	78&	86&	76\\
        \hline

        13 & 80&	82&	87&	76&	83&	82	&91	&86&	84&	84\\
        \hline

        14 & 82	&83	&88&	77&	85&	78	&92	&75	&85	&80\\
        \hline

        15 & 80&	82&	87&	86&	83&	82&	91&	76&	84	&74\\
        \hline

        Average	&78.6&	79.7	&84.7	&78.6&	80.8&	77.6	&88.7&	80.8&	82.3	&83.3\\
        \hline

        Standardization&	0.19	&0.31&	0.83	&0.19	&0.42	&0.13	&0.98	&0.42&	0.59	&0.70\\
        \hline

        Weight of $A_{i,j}^r$&	0.04	&0.065	&0.175&	0.04&	0.088	&0.027&	0.206&	0.088&	0.124	&0.147\\
        \hline
    \end{tabular}}
    
    \label{tab3}
\end{table*}

\emph{1) Memory Ability:} After 20 rounds of interaction, each LLM is prompted to recall and restate the initial questions and data presented in the first round. This test evaluates the model’s ability to retain and recall historical information accurately, which is essential to maintain contextual consistency over extended dialogues.

\emph{2) Daily Life Ability:} A set of practical, real-world tasks is designed, including weather forecasting, fraud detection, cooking advice, and other everyday scenarios. This evaluation assesses each LLM's grasp of common-sense knowledge and its capacity to assist users in routine daily activities.

\emph{3) Artistic Ability:} LLMs are prompted to engage in creative tasks such as composing poetry, generating aesthetic commentary, producing musical pieces, and simulating traditional forms of art like Chinese ink painting. The goal is to assess the models' expressive and creative capabilities in supporting human artistic endeavors.

\emph{4) Logical Reasoning Ability:} This dimension involves mathematical problem-solving, optimization tasks relevant to network systems, and performance analysis in wireless communication. It is designed to measure the logical reasoning and analytical capabilities of each LLM in handling structured, domain-specific challenges.

\emph{5) Code Generation Ability:} Building on the logical reasoning tasks, each LLM is asked to generate implementation code in various programming languages, including MATLAB, Python, and C++. The results are compared to evaluate the LLM's effectiveness in translating abstract problem-solving into executable solutions, demonstrating their potential in software development contexts.

Table~\ref{tab3} presents the average scores assigned by 15 volunteers to 10 representative LLMs across five distinct evaluation dimensions. Each dimension is scored on a scale from 0 to 100. Following score aggregation, a standardization process is applied to prepare the data for subsequent analysis. Specifically: \emph{a)} the raw scores are transformed into a standard normal distribution with a mean of 0 and a standard deviation of 1 (i.e., Z-distribution); and \emph{b)} a cumulative distribution function is used to map the standardized values into a uniform distribution over the interval [0, 1]. Based on these normalized values, the quality weights of the LLM responses are calculated by Eq.~(\ref{eq2}).

It is important to note that we do not focus on modeling the trustworthiness of LLMs, as extensive previous work has addressed trust and reputation assessment schemes for network nodes \cite{guo2024trusted}, \cite{liao2024graph}. In our framework, trust is characterized as the probability of malicious behavior exhibited by an LLM. We assume that trust weights follow a normal distribution, such as $N(0.1, 0.2)$, or another suitable distribution, with values constrained to the interval (0, 1).

\subsection{Consensus Performance}
In this part, we compare the security, latency, and throughput of WBFT against three established consensus mechanisms, i.e., PBFT \cite{castro1999practical}, Vote as Proof (VaaP) \cite{fu2022votes}, and Artificial Bee Colony-PBFT (ABC-PBFT) \cite{xu2023abc}, under varying proportions of response quality and trust weights. 
Specifically, we evaluate three configurations: $\alpha = 0.4$, $\beta = 0.6$; $\alpha = 0.5$, $\beta = 0.5$; and $\alpha = 0.6$, $\beta = 0.4$. 
These comparison baselines are widely adopted as effective defenses against Byzantine attacks and represent three distinct categories of Byzantine Fault Tolerance (BFT): classical BFT (PBFT), parallel BFT (VaaP), and BFT with selected reliable nodes (ABC-PBFT).

In the WBFT and ABC-PBFT schemes, five LLMs are selected as consensus nodes using the HSC and ABC algorithms, respectively. 
Consensus security reflects the robustness of MultiLLMN in the presence of malicious LLMs, while latency and throughput capture the system's efficiency in responding to UE requests. 
To evaluate performance under different trust conditions, we construct three trust weight distributions for the 10 LLMs, i.e., $N(0.1, 0.6)$, $N(0.1, 0.4)$, and $N(0.1, 0.2)$. 
The mean is kept constant at 0.1 to ensure that the total trust weight across all LLMs remains normalized to 1.

\begin{figure*}[!t]
   \centering
  \includegraphics[width=7.2in]{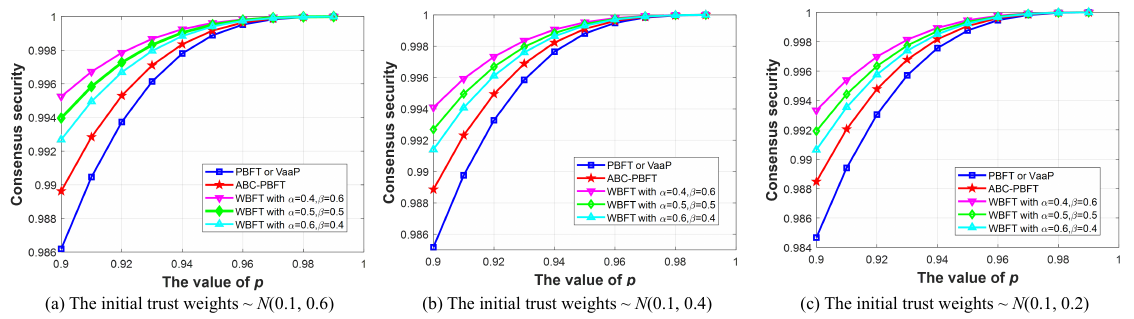}
   \caption{Consensus security. (a) The initial trust weights of LLMs follow a $N(0.1,0.6)$ distribution. (b) The initial trust weights of LLMs follow a $N(0.1,0.4)$ distribution. (c) The initial trust weights of LLMs follow a $N(0.1,0.2)$ distribution.}
   \label{fig4}
\end{figure*}

\begin{figure*}[!t]
   \centering
  \includegraphics[width=7.2in]{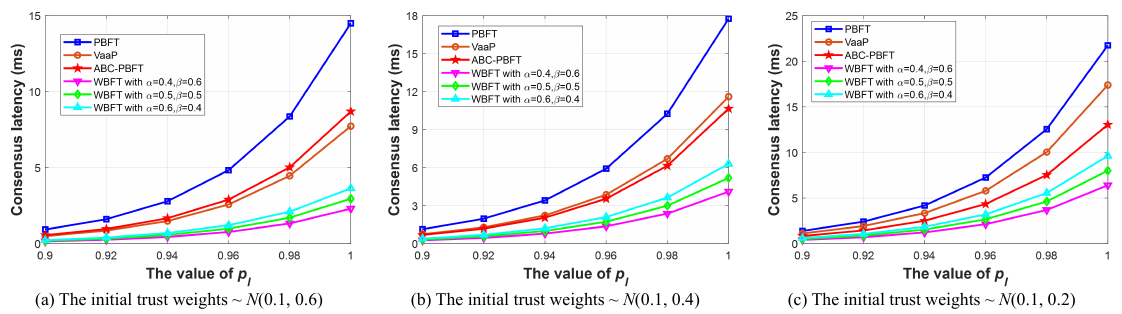}
   \caption{Consensus latency. (a) The initial trust weights of LLMs follow a $N(0.1,0.6)$ distribution. (b) The initial trust weights of LLMs follow a $N(0.1,0.4)$ distribution. (c) The initial trust weights of LLMs follow a $N(0.1,0.2)$ distribution.}
   \label{fig5}
\end{figure*}

\emph{1) Consensus Security}: 
Figs. \ref{fig4} (a)-(c) show that WBFT has significantly better consensus security than other consensus. This is because WBFT effectively reduces the impact of Byzantine attacks by suppressing the voting power of malicious nodes. PBFT and VaaP result in the same performance due to a fixed fault tolerance threshold $2/3$. Although ABC-PBFT benefits from selecting some reliable LLMs as consensus nodes, it has higher consensus security than these two. However, the consensus nodes it selects are static and cannot be dynamically adjusted according to the network conditions and LLM trust, compared to WBFT. Therefore, its consensus security is lower than that of WBFT.

Additionally, in various combinations of weight ratios, the trust weight is of greater importance to consensus security than the response quality weight. The reason for this is that when the value $\beta$ is high, the WBFT consensus is better able to exert the voting power of a trusted LLM. 

Furthermore, we observe that as the variance $\sigma^2$ of the initial trust weight distribution $B_{i,j}^r$ decreases, the consensus security of WBFT also drops. This phenomenon can be attributed to the fact that variance reflects the dispersion of trust weights across LLMs. A higher variance implies that some LLMs may receive significantly higher trust weights, increasing their influence in the consensus process and thereby enhancing the likelihood of achieving consistency. Mathematically, as shown in Eq.~(\ref{eq6}), the mean trust value remains fixed at 0.1, which is below the Byzantine fault tolerance threshold of $2/3$. Taking the derivative of the consensus probability with respect to the variance yields a positive value, indicating that an increase in variance shifts the distribution rightward and raises the probability of surpassing the consensus threshold. Additionally, we can observe that WBFT exhibits greater improvements in consensus security compared to the other two baselines when the trust weight variance is high. This suggests that the influence of highly trusted LLMs is more effectively leveraged in WBFT under conditions of greater trust dispersion.

\emph{2) Consensus Latency}: 
To evaluate consensus latency, we set the bandwidth $B$ to 15 kHz, the channel capacity $C$ to 15 kbps, and the transmission rate $R$ to 10 kbps, respectively. We suppose that the MultiLLMN has 4 clusters. Then, we use the probability $P_l$ as a horizontal coordinate to study and compare the latency of each consensus and its influence parameters.

Figs. \ref{fig5} (a)-(c) respectively illustrate the latency of WBFT and other baselines under different initial trust weight distributions. In either case, WBFT achieves lower latency than other consensus, showing that it can efficiently respond to UE needs and provide trusted answers. This is because the two-round voting consensus and our pipeline mechanism improve consensus efficiency. Meanwhile, compared with VaaP, WBFT has a higher consensus security, thus avoiding as much as possible the delay of leader re-election. Importantly, clustering also shrinks the scope of consensus voting, thereby further reducing consensus latency.

Additionally, the ratio of response quality and trust weights also affects consensus latency by affecting consensus security. Specifically, the experimental results demonstrate that latency is further reduced with increasing $\beta$ value, primarily due to the positive correlation between increased trust weight and increased probability of consensus success.  It effectively mitigates the occurrence of re-election scenarios where a leader struggles to secure the required $2/3$ vote weight threshold.
Moreover, the variance $\sigma^2$ of the initial trust weight distribution $B_{i,j}^r$ impacts consensus latency as well. Specifically, as the variance decreases, the latency increases. This is because lower variance reduces consensus security, leading to a higher likelihood of leader re-elections and consequently prolonging the consensus process. We further observe that under high-variance conditions, WBFT achieves significantly lower latency compared to the other consensus mechanisms, which can be attributed to its stronger consensus security, as previously discussed.

An additional observation concerns the performance of ABC-PBFT and VaaP under varying trust weight variances. When $\sigma^2$ is large, ABC-PBFT benefits from its ability to pre-select reliable LLMs, enabling faster consensus and outperforming VaaP in terms of efficiency. However, when $\sigma^2$ is small and trust differences among LLMs become less pronounced, ABC-PBFT loses its advantage, and its latency performance converges with or falls behind that of VaaP. This further highlights the effectiveness of the HSC algorithm used in WBFT, which demonstrates superior capability in identifying trusted, low-latency consensus nodes compared to the ABC-based approach.

\begin{figure*}[!t]
   \centering
  \includegraphics[width=7.2in]{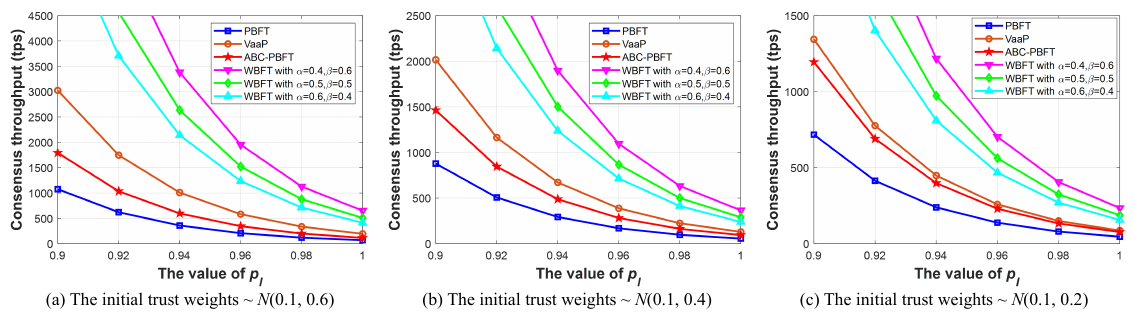}
   \caption{Consensus throughput. (a) The initial trust weights of LLMs follow a $N(0.1,0.6)$ distribution. (b) The initial trust weights of LLMs follow a $N(0.1,0.4)$ distribution. (c) The initial trust weights of LLMs follow a $N(0.1,0.2)$ distribution.}
   \label{fig6}
\end{figure*}

\begin{figure*}[!t]
   \centering
  \includegraphics[width=7.2in]{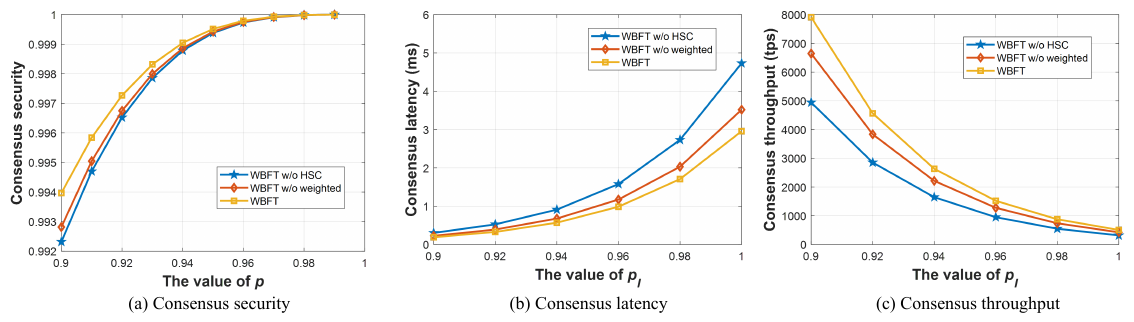}
   \caption{Ablation test of WBFT when the initial trust weights follow a $N(0.1,0.6)$ with $\alpha=\beta=0.5$. (a) Consensus security. (b) Consensus latency. (c) Consensus throughput.}
   \label{fig7}
\end{figure*}

\emph{3) Consensus Throughput}:
Consensus throughput is evaluated under the same simulation environment and configuration settings as those used for consensus latency, with results presented in Figs.~\ref{fig6}(a)–(c).

In general, throughput exhibits an inverse relationship to latency. Nevertheless, consistent with latency results, WBFT consistently demonstrates the highest throughput performance among all baselines. The reasons are twofold. First, WBFT employs a multi-chain and pipelined architecture, enabling it to handle multiple UE requests concurrently. Second, its high consensus security reduces the likelihood of disruptions caused by malicious LLMs. Furthermore, the proposed HSC algorithm effectively narrows the consensus scope, further contributing to improved throughput.

Additionally, the ratio of response quality to trust weights, as well as the initial trust weight distribution variance $\sigma^2$ of LLMs, exhibit effects on throughput that are consistent with their impacts on consensus security and latency. For VaaP and ABC-PBFT, variations in trust weight variance $\sigma^2$ do not alter the relative throughput performance between the two schemes, unlike their impact on latency. VaaP consistently achieves higher throughput than ABC-PBFT, primarily due to its parallel block processing mechanism.

\emph{4) Ablation Study:}
Through comparative experiments, we have demonstrated that WBFT achieves superior consensus security and efficiency compared to existing consensus mechanisms. To further validate the contribution of individual modules within WBFT, we conduct ablation tests focusing on two key components: the HSC algorithm and the weight-allocation scheme. The simulation parameters for these tests are consistent with those used in earlier consensus performance evaluations.

The results of the ablation study in terms of consensus security, latency, and throughput are illustrated in Figs.~\ref{fig7}(a)–(c). Note that the configuration without the HSC algorithm is denoted as w/o HSC, and the configuration without the weight-allocation scheme is denoted as w/o Weighted. These results are obtained under the conditions where the initial trust weights follow $N(0.1, 0.6)$ and $\alpha = \beta = 0.5$.
From Fig.~\ref{fig7}, we can observe that the absence of either the HSC algorithm or the weight allocation mechanism leads to degraded consensus performance. Specifically, the HSC algorithm enhances consensus by excluding LLMs with low trust values or high communication latency, thereby reducing the influence of malicious LLMs and improving overall consensus efficiency. Meanwhile, the weight-allocation scheme decreases the voting influence of less reputable LLMs, further strengthening consensus security. It also increases the consensus success rate and reduces latency by minimizing leader reselection events caused by consensus failures. In conclusion, these findings underscore the indispensable role of the HSC algorithm and weight allocation scheme in achieving the performance improvements observed with WBFT.
\begin{figure}[!t]
\centering
 \includegraphics[width=3.4in]{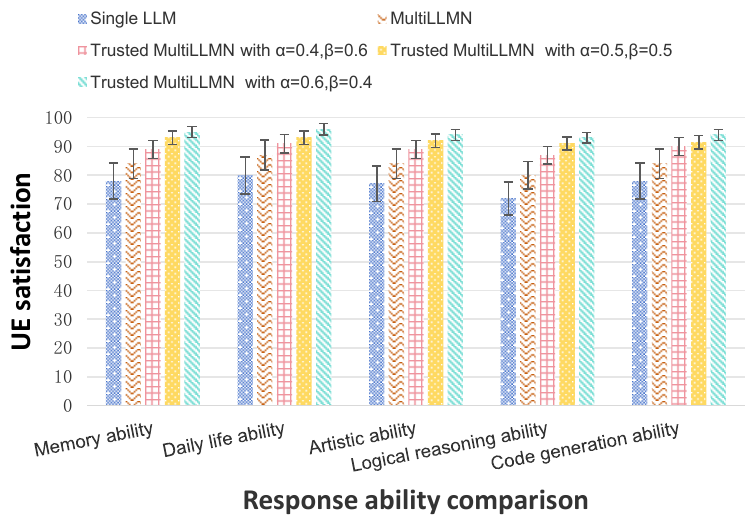}
   \caption{UE satisfaction comparison.}
\label{fig8}
\end{figure}

\subsection{UE Satisfaction}
In this part, we evaluate UE satisfaction with three different architectures: a single LLM, a MultiLLMN without blockchain consensus, and the Trusted MultiLLMN proposed in this work. The evaluation is carried out using the set of questions described in Section~\ref{sec-VI}-A. UE satisfaction is synthesized by surveying the same group of 15 volunteers, each scoring responses on a 100-point scale.
Specifically, the single LLM scenario involves volunteers randomly using one of the LLMs evaluated in Section~\ref{sec-VI}-A. The MultiLLMN configuration refers to a centralized architecture without blockchain consensus participation, leaving the system vulnerable to the influence of malicious LLMs. In contrast, the Trusted MultiLLMN represents the WBFT consensus-driven decentralized system we have designed. The comparison among these three architectures essentially serves as an ablation study to assess the individual and combined contributions of LLM collaboration and WBFT consensus to overall generation quality.

Fig.~\ref{fig8} illustrates the UE satisfaction of three architectures across five evaluation dimensions. Notably, the single LLM setup yields the lowest satisfaction scores and the highest standard deviation, highlighting its limited ability to generalize across diverse tasks and the inherent variability among different LLMs. In comparison, WBFT-driven Trusted MultiLLMN achieves higher average satisfaction and reduced variability compared to baseline MultiLLMN. This improvement is attributed to the consensus mechanism, which mitigates the risk of misinformation from Byzantine LLMs and strengthens overall system reliability.
Furthermore, increasing the weight assigned to response quality in the WBFT consensus process is found to correlate with higher UE satisfaction. However, this improvement in perceived quality comes at the expense of reduced consensus efficiency, suggesting a trade-off between optimizing consensus performance and maximizing user satisfaction. This observation highlights an important design consideration for deploying Trusted MultiLLMN architectures in future applications.

\section{Conclusion} \label{sec-VII}
In this paper, we have proposed a MultiLLMN framework to address the inherent limitations of individual LLMs, such as hallucinations, biased outputs, and limited generalization caused by static or incomplete training data. To ensure that collaboration among multiple LLMs produces reliable and trustworthy responses, we have presented the Trusted MultiLLMN driven by blockchain. Moreover, we have proposed the WBFT consensus mechanism. Unlike traditional consensus methods that assign equal voting rights or rely on fixed trust assumptions, WBFT dynamically integrates both response quality and trust levels to guide consensus decisions. Our simulation results have demonstrated that WBFT not only enhances consensus security in the presence of potentially malicious LLMs but also significantly reduces latency and increases throughput compared to PBFT, VaaP, and ABC-PBFT. Furthermore, the Trusted MultiLLMN built on WBFT exhibits stronger responsiveness and reliability than both standalone LLMs and collaborative LLM structures that lack consensus coordination.



\bibliographystyle{IEEEtran}
\bibliography{IEEEabrv,mylib}

\newpage


\vspace{11pt}

\vfill

\end{document}